\documentclass[3p,times]{elsarticle}

\usepackage{ecrc}

\volume{00}

\firstpage{1}

\journalname{Physics Letters A}

\runauth{}

\jid{pla}

\jnltitlelogo{Physics Letters A}

\usepackage{amssymb}
\usepackage{graphicx}
\usepackage{color}
\usepackage{url}

\biboptions{compress}

\usepackage[figuresright]{rotating}

\begin{document}

\begin{frontmatter}



\dochead{}
%% e.g. \dochead{17th International Conference on Dynamical Processes in Excited States of Solids}

\title{String networks in $Z_N$ Lotka-Volterra competition models}


\author[cfa-up,dfa-up]{P. P. Avelino}
%\email[Electronic address: ]{Pedro.Avelino@astro.up.pt}
\address[cfa-up]{Centro de Astrof\'{\i}sica da Universidade do Porto, Rua das Estrelas, 4150-762 Porto, Portugal}
\address[dfa-up]{Departamento de F\'{\i}sica e Astronomia, Faculdade de Ci\^encias, Universidade do Porto, Rua do Campo Alegre 687, 4169-007 Porto, Portugal}
\ead{Pedro.Avelino@astro.up.pt}
\ead{Tel: +351226089843}
\ead{Fax: +351226089831}

\author[if-usp,df-ufpb]{D. Bazeia}
%\email[Electronic address: ]{bazeia@fisica.ufpb.br}
\address[if-usp]{Instituto de F\'\i sica, Universidade de S\~ao Paulo, 05314-970 S\~ao Paulo, SP, Brazil}
\address[df-ufpb]{Departamento de F\'{\i}sica, Universidade Federal da Para\'{\i}ba, 58051-970 Jo\~ao Pessoa, PB, Brazil}

\author[cfp-up,ect-ufrn]{J. Menezes}  
%\email[Electronic address: ]{jmenezes@ect.ufrn.br} 
\address[cfp-up]{Centro de F\'{\i}sica do Porto, Rua do Campo Alegre 687, 4169-007 Porto, Portugal} 
\address[ect-ufrn]{Escola de Ci\^encias e Tecnologia, Universidade Federal do Rio Grande do Norte\\ Caixa Postal 1524, 59072-970 Natal, RN, Brazil}

\author[dfi-uem]{B. F. de Oliveira}  
%\email[Electronic address: ]{breno@dfi.uem.br} 
\address[dfi-uem]{Departamento de F\'{\i}sica, Universidade Estadual de Maring\'a, Av. Colombo, 5790, 87020-900 Maring\'a, PR, Brazil}

\begin{abstract}
In this letter we give specific examples of $Z_N$ Lotka-Volterra competition 
models leading to the formation of string networks. We show that, in order 
to promote coexistence, the species may arrange themselves around regions 
with a high number density of empty sites generated by predator-prey 
interactions between competing species. These configurations extend 
into the third dimension giving rise to string networks. We investigate 
the corresponding dynamics using both stochastic and mean field theory 
simulations, showing that the coarsening of these string networks 
follows a scaling law which is analogous to that found in other 
physical systems in condensed matter and cosmology.
\end{abstract}

\begin{keyword}
May-Leonard models \sep string networks 



\end{keyword}

\end{frontmatter}

% \linenumbers

\section{Introduction}
\label{introduction}

May-Leonard models, also known as rock-paper-scissors type models,
incorporate important rules associated with the dynamics
of networks of competing populations and are widely regarded as a 
useful tool in the understanding of the mechanisms leading to
biodiversity \cite{May-Leonard, Kerr2002, Reichenbach2007}  (see also \cite{Volterra,doi:10.1021/ja01453a010} for the pioneer work by Lotka 
and Volterra). The simplest models usually consider three species and 
allow for three basic actions (motion, reproduction and predation) but 
several generalizations incorporating other interactions and further species 
have also been proposed in the literature \cite{Szabo2008, Peltomaki2008, 
Durrett:2009, PhysRevE.81.046113, Edwards:2010, PhysRevE.85.061924, 
Kang20132652, Lutz2013286}.

The study of the spatial dynamics of stochastic May-Leonard models in
two-dimensions has revealed \cite{PhysRevLett.99.238105,
PhysRevE.83.011917, May-Leonard-Mobilia, PhysRevE.86.021911,
PhysRevE.86.031119, PhysRevE.86.036112, PhysRevE.87.032148,
1742-5468-2012-07-P07014, Jiang20122292} the emergence of complex
spiralling patterns and interfaces (with or without junctions) whose
dynamics is controlled by the strength of the interactions between the
individuals of the various species. In some of the models, the interface
dynamics has been shown to be curvature driven, in close parallel with
the dynamics of cosmological domain walls \cite{Avelino2005,
Avelino2008, PhysRevD.79.085007} or non-relativistic interfaces in
condensed matter systems \cite{Stavans1989, Glazier1992, Flyvbjerg1993,
Monnereau1998, Weaire2000, PhysRevE.74.061605}.

Interfaces in ideal soap froths and grain growth are characterized by an 
interface velocity which is proportional to the mean curvature at each 
point. This property naturally leads to an increase of the characteristic 
scale of the network $L$ with physical time $t$ as $L \propto t^{1/2}$, as 
a result of the gradual elimination of entire domains. On the other hand, the dynamics 
of cosmological domain wall, string and other p-brane networks has also been 
investigated in detail due to their potential observational signatures. In particular, 
it has been demonstrated \cite{Avelino:2010qf, Avelino:2011ev, Sousa:2011ew, 
Sousa:2011iu} that the same phase field and velocity dependent 
one-scale models characterizing the dynamics of relativistic p-brane networks, 
in a cosmological context, can also successfully describe, in a friction 
dominated regime, the dynamics of interfaces and strings in a wide variety 
of material of condensed matter systems (with the $L \propto t^{1/2}$ scaling regime being 
obtained for a homogeneous and time independent friction lengthscale \cite{PhysRevD.53.R575}).

In the present letter we show that string networks may arise in the
context of spatial stochastic models in three spatial dimensions.
We study specific models where predator-prey interactions give rise to
strings characterized by core regions with a high density of empty
sites, investigating the corresponding evolution using both stochastic
and mean field theory simulations. We determine the scaling laws
governing the string network dynamics, comparing with the corresponding
two-dimensional evolution and with curvature driven string network
dynamics in other physical systems.

\section{The model}
\label{the_model}

We consider a $Z_N$ Lotka-Volterra sub-family of the more general family of 
spatial stochastic May-Leonard models introduced in Refs. 
\cite{PhysRevE.86.031119,PhysRevE.86.036112}. We focus on models with symmetric predator-prey 
interactions in which an individual of any of the $N$ species predates and is 
hunted by individuals of $N-3$ other species, with probability $p$. In these 
models individuals of $N$ species and some empty sites (E) are initially 
distributed on a square lattice with ${\mathcal N}$ sites (the grid 
spacing is $\Delta x=1$). The different species are
labelled by $i$ (or $j$) with $i, j= 1,\ldots,N$, and the cyclic
identification $i= i+ k\, N$ where $k$ is an integer, is made. The sum
of the number of individuals of the species $i$ ($I_i$) and empty sites
($I_E$) is equal to the number of sites (${\mathcal N}$). At each time
step a random individual (active) interacts with one of its four nearest
neighbors (passive). The unit of time $\Delta t=1$ is defined as the
time necessary for ${\mathcal N}$ interactions to occur (one generation
time). The possible interactions are classified as Motion $ i\ \odot \to
\odot\ i\,, $ Reproduction $ i\ \otimes \to ii\,, $ or Predation $i\
\ (i+\alpha) \to i\ \otimes\,, $ where $\odot$ may be any species ($i$)
or an empty site ($\otimes$) and $\alpha=2,\ldots,N-2$.

\begin{figure}[h]
	\centering
	\includegraphics[scale=0.7]{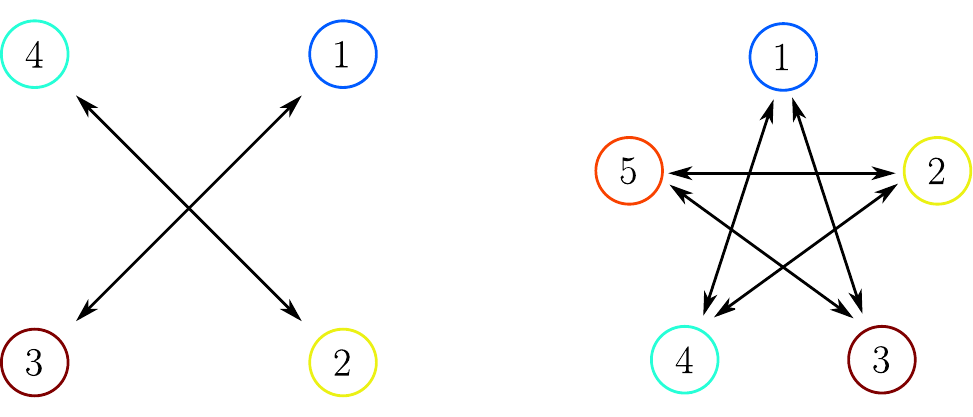}
	\caption{Possible predator-prey interactions in the cases with $4$ and $5$ species.}
	\label{fig-1}
\end{figure}

Fig. \ref{fig-1} illustrates the possible predator-prey interactions in the
cases with $4$ and $5$ species. For simplicity, we also assume that the
Motion ($m$) and Reproduction ($r$) probabilities are the same for all
species. Throughout this letter we shall take $m=0.15$, $r=0.10$ and
$p=0.75$. However, we verified that our main results also hold for other
choices of $m$, $r$ and $p$.
%%%%%%%%%%%%%%%%%%%%%%%%%%%%%%%%%%%%%%%%%%%%%%%%%%%%%%%%%%%%
%%%%%%%%%%%%%%%%%%%%%%%%%%%%%%%%%%%%%%%%%%%%%%%%%%%%%%%%%%%%

As soon as the simulations start, individuals of the same species,
originally spread out randomly throughout the lattice tend to join each
other to share common spatial regions. Moreover these regions are
arranged so that they are bounded by other domains dominated by species
with which there is no predator-prey interaction. Such spatial
configurations promote the coexistence of cooperating domains which may
organize, either clockwise or counterclockwise, around core regions with
a significantly higher density of empty sites. We define these configurations 
as defect or anti-defects depending on whether the species $i, i+1, ..., i+N-1$ are arranged 
clockwise or counterclockwise, respectively (string-like defects and 
anti-defects are characterized by symmetric winding numbers associated 
to the clockwise or anti-clockwise vortex states). In two spatial dimensions, these are roughly
circular regions, created by predator-prey interactions between
competing species. They define the defect/anti-defect cores whose
average area is a function of the interaction probabilities ($m$, $r$
and $p$) but is roughly constant in time. When a defect is close 
to an anti-defect, they may form a defect/anti-defect pair which is 
unstable to collapse (see, for example, \cite{Vilenkin:2000} for more details).

\begin{figure}[h]
	\centering
	\includegraphics[scale=0.7]{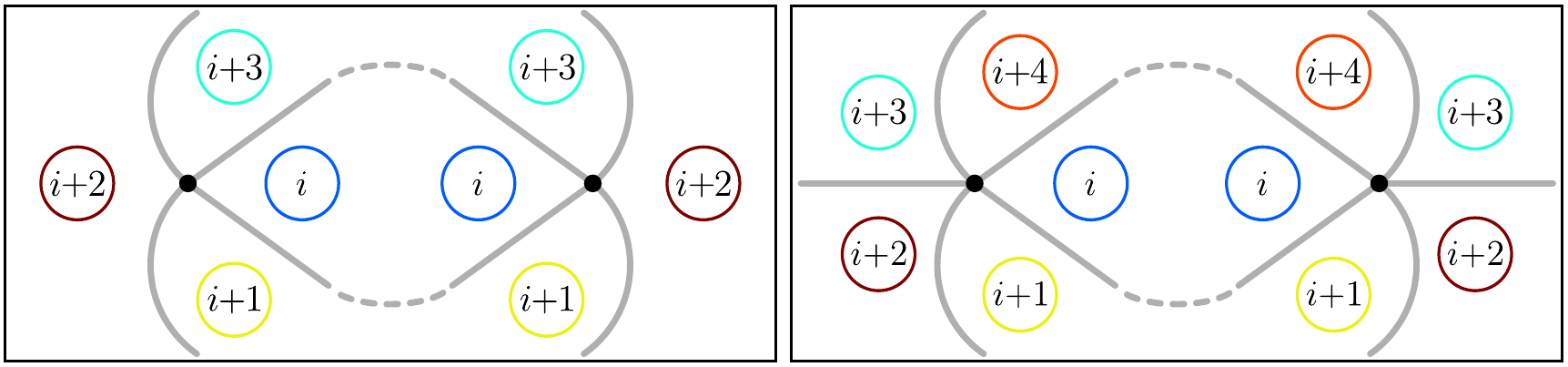}
	\caption{This figure schematically describes the 
	organization of the species around the defect cores. 
	The black dots represent a topological defect/antidefect pair 
	of 4 (left panel) and 5 (right panel) species.}
	\label{fig-2}
\end{figure}

Consider the defect/anti-defect pair of $4$ species shown in the left 
panel of Fig. \ref{fig-2}. The average number of attacks per unit time
from individuals of the species $i + 2$ is larger than the average
number of attacks from individuals of the inner species $i$. This
implies that individuals of the outer species tend to invade the
territory of the inner ones causing an approximation and annihilation of
the defect/anti-defect pair. We shall show that the defect/anti-defect
cores attract each other, having a velocity which is, on average,
inversely proportional to the distance between them. In contrast, a pair
of clockwise (or counterclockwise) defects (anti-defects) cannot annihilate; on the
contrary, they repel each other. The analogous $5$ species case is shown
in the right panel of Fig. \ref{fig-2}.
%%%%%%%%%%%%%%%%%%%%%%%%%%%%%%%%%%%%%%%%%%%%%%%%%%%%%%%%%%%%
%%%%%%%%%%%%%%%%%%%%%%%%%%%%%%%%%%%%%%%%%%%%%%%%%%%%%%%%%%%%

\section{Results}
\label{results}

\begin{figure}[h]
	\centering
	\includegraphics[width=4.0cm]{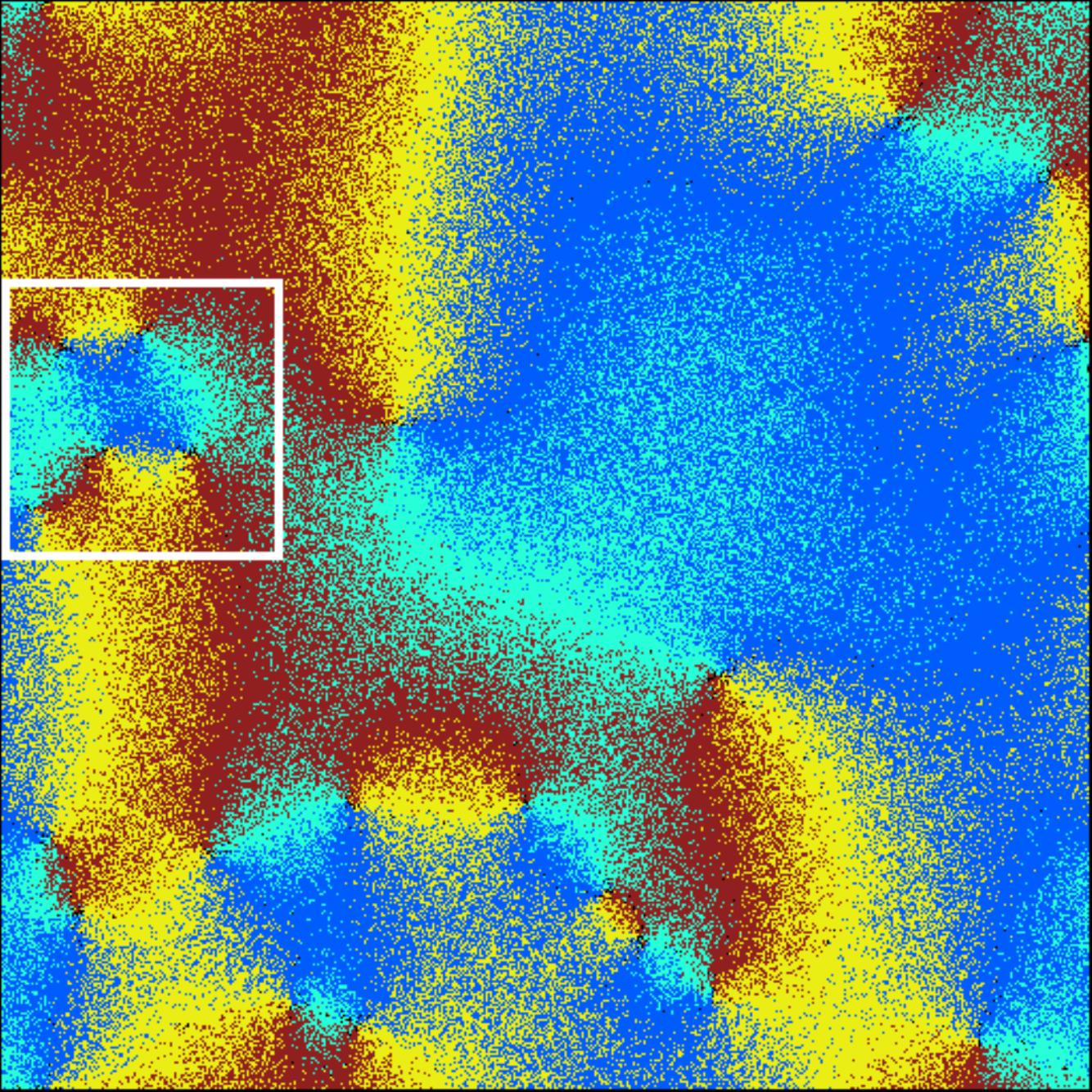}
		\hspace{0.1cm}
	\includegraphics[width=4.0cm]{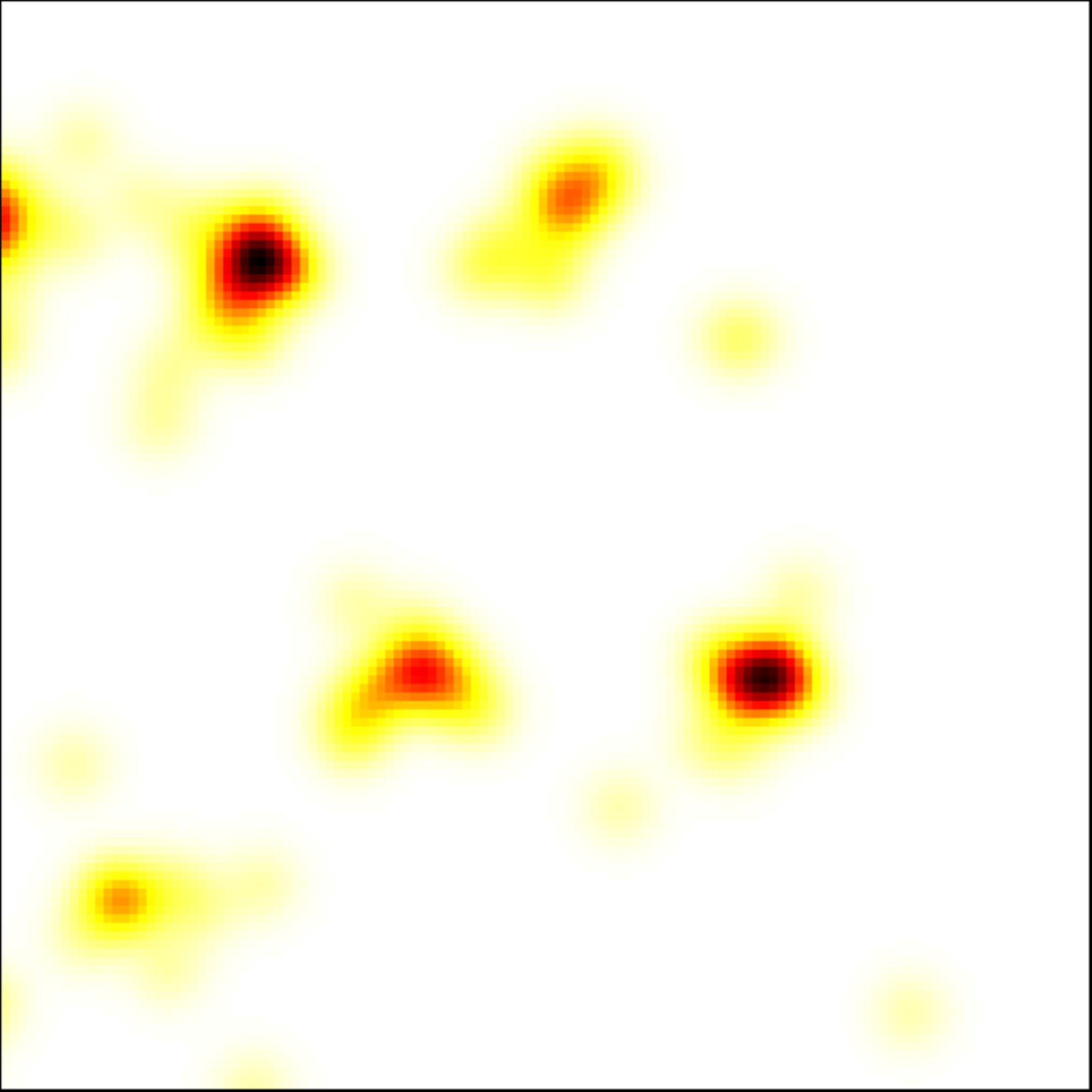}
\\ \vspace{0.2cm}
	\includegraphics[width=4.0cm]{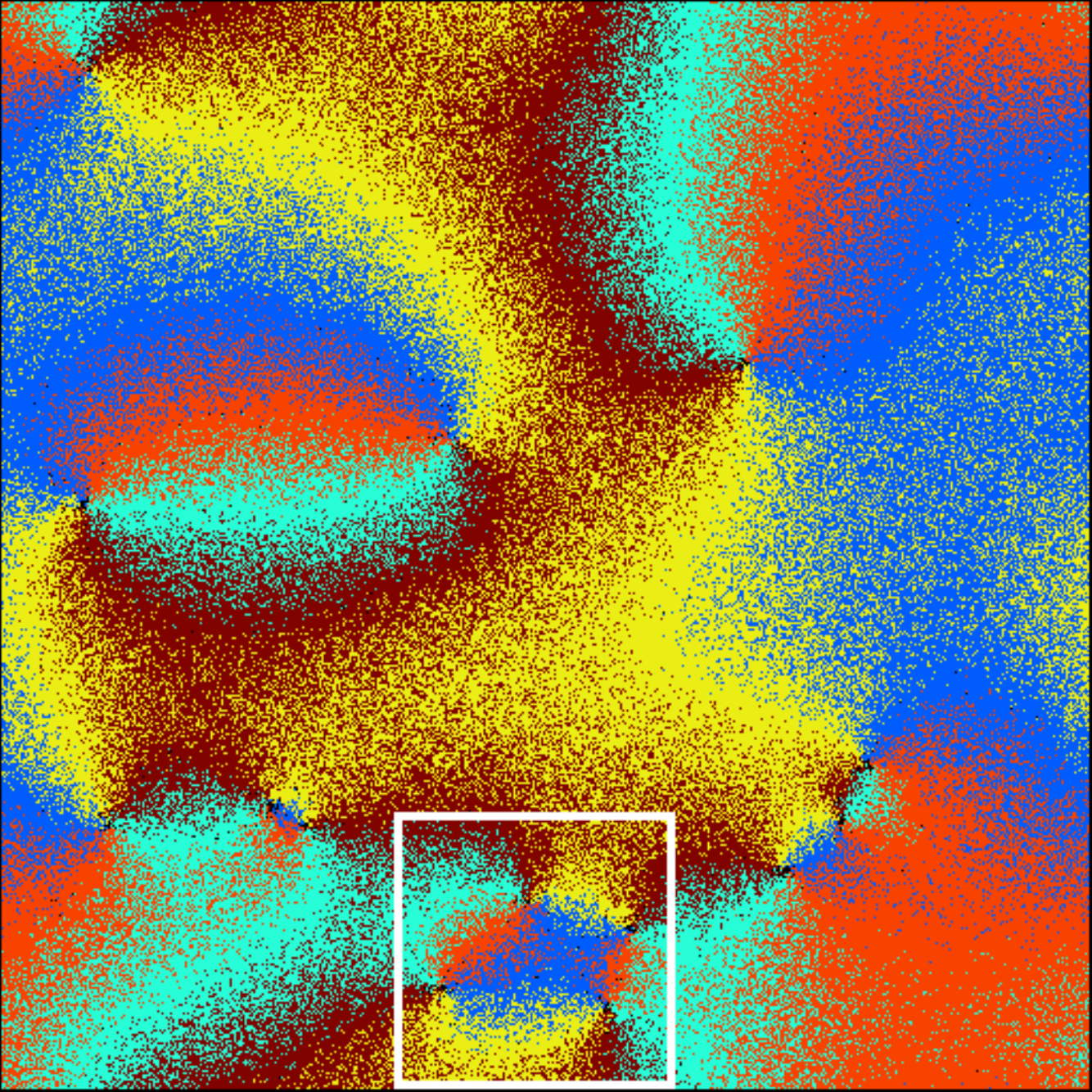}
		\hspace{0.1cm}
	\includegraphics[width=4.0cm]{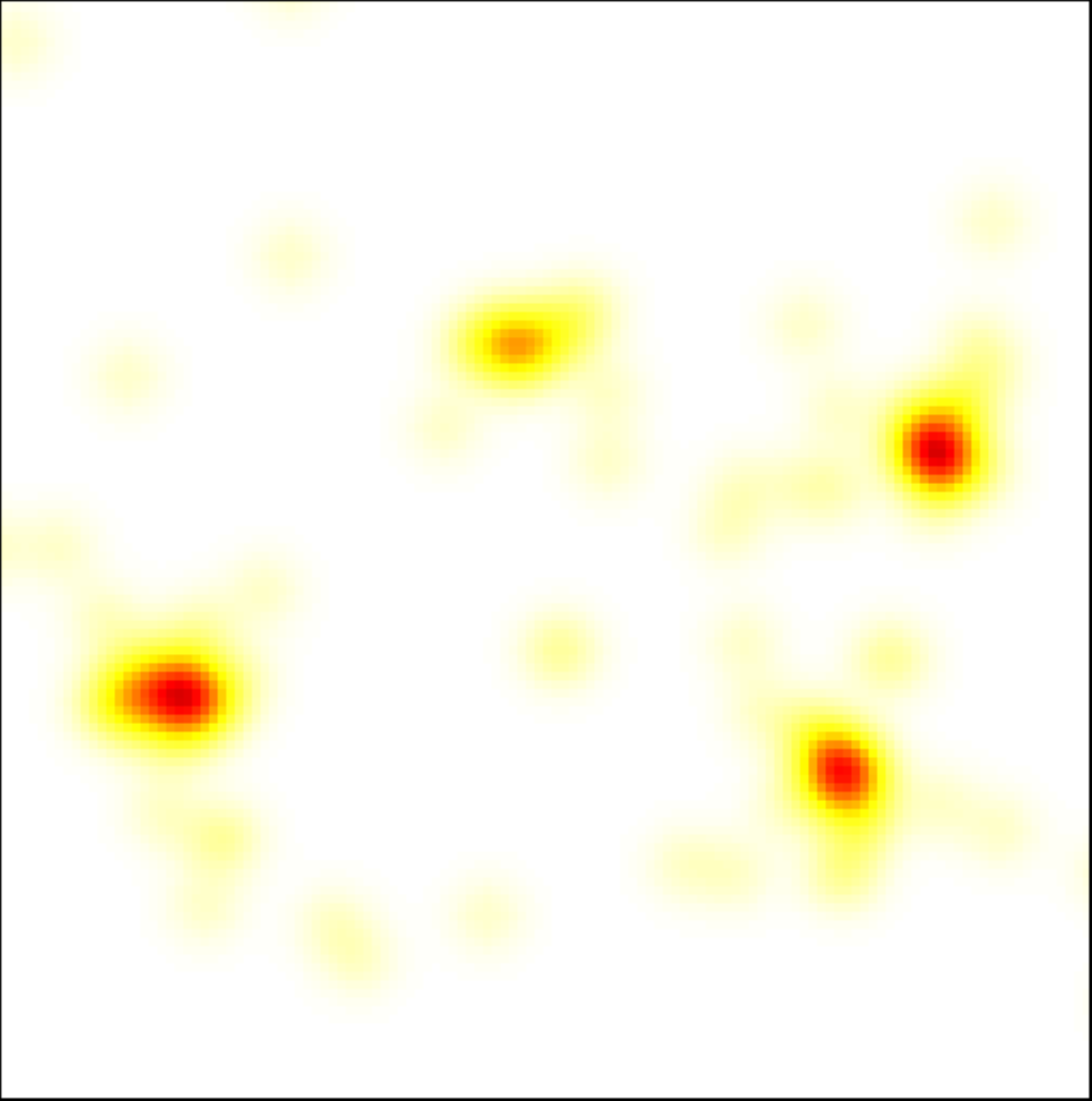}
\caption{Snapshot of the evolution of the $4$ (upper panels) and $5$ (bottom
panels) species model on a $512^2$ lattice after $5000$ generations. The core 
regions in the right panels represent the defect cores inside the
small region of the left panel indicated by a box.}
\label{fig-3}
\end{figure}

Figure \ref{fig-3} shows two snapshots (taken after $5000$ generations) from two-dimensional $512^2$ simulations, with periodic boundary conditions,
of the $4$ species model (upper panels) and the $5$ species model (bottom panels). The colors blue, yellow, brown, light blue and
orange in the left panel represent the species 1, 2, 3, 4 and 5
respectively. One sees that the individuals of the various species
dispose themselves around core regions where most of the predator-prey
interactions take place. Note that, as expected given the periodic
boundary conditions, there is an equal number of defects and anti-defects 
present. The empty sites, concentrated mainly at
the defect/anti-defect cores, are highlighted in the darker regions of the right
panel, showing two pairs of defects/anti-defects. As the network evolves, 
the number of empty sites decreases due to the annihilation of defect/anti-defect pairs,  
analogous to the coarsening dynamics of liquid crystal textures in two-dimensional 
nematic liquid crystals \cite{Oliveira:2010}.

%%%%%%%%%%%%%%%%%%%%%%%%%%%%%%%%%%%%%%%%%%%%%%%%%%%%%%%%%%%%
%%%%%%%%%%%%%%%%%%%%%%%%%%%%%%%%%%%%%%%%%%%%%%%%%%%%%%%%%%%%

\begin{figure}[h]
	\centering
	\includegraphics[width=4.0cm]{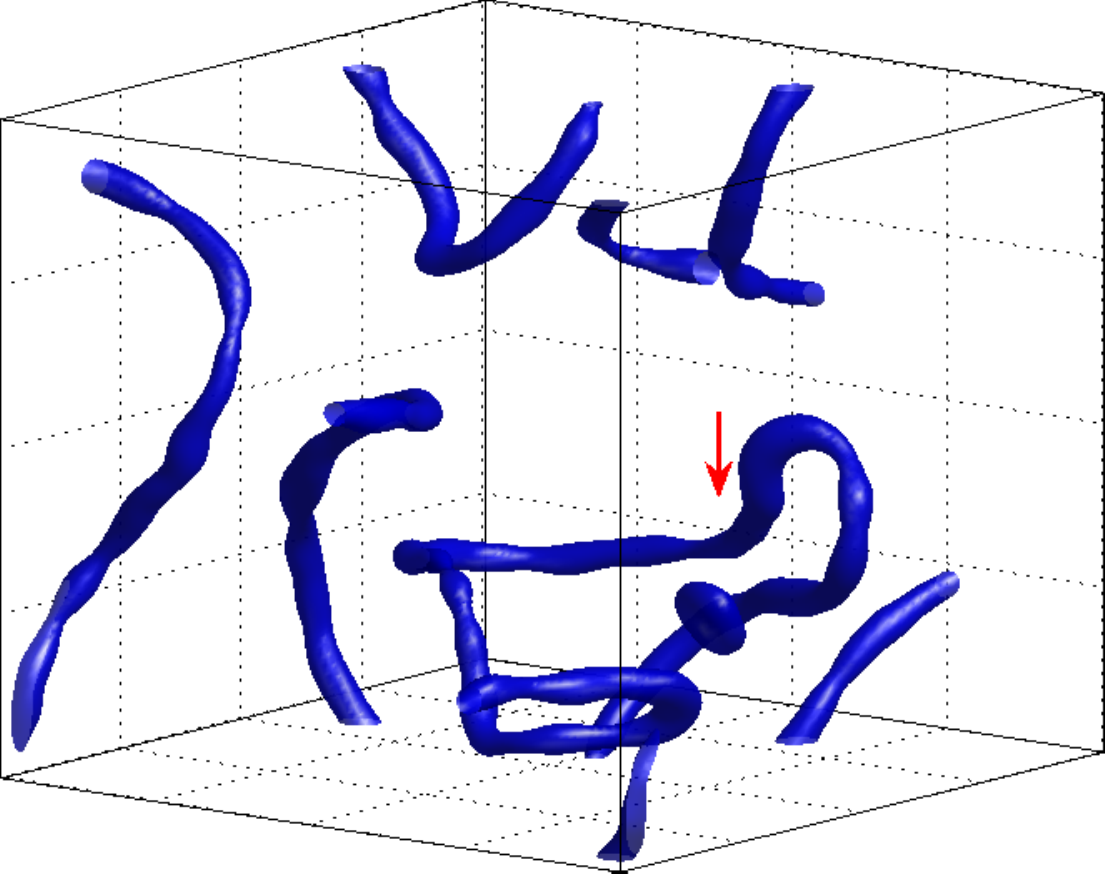}
	\includegraphics[width=4.0cm]{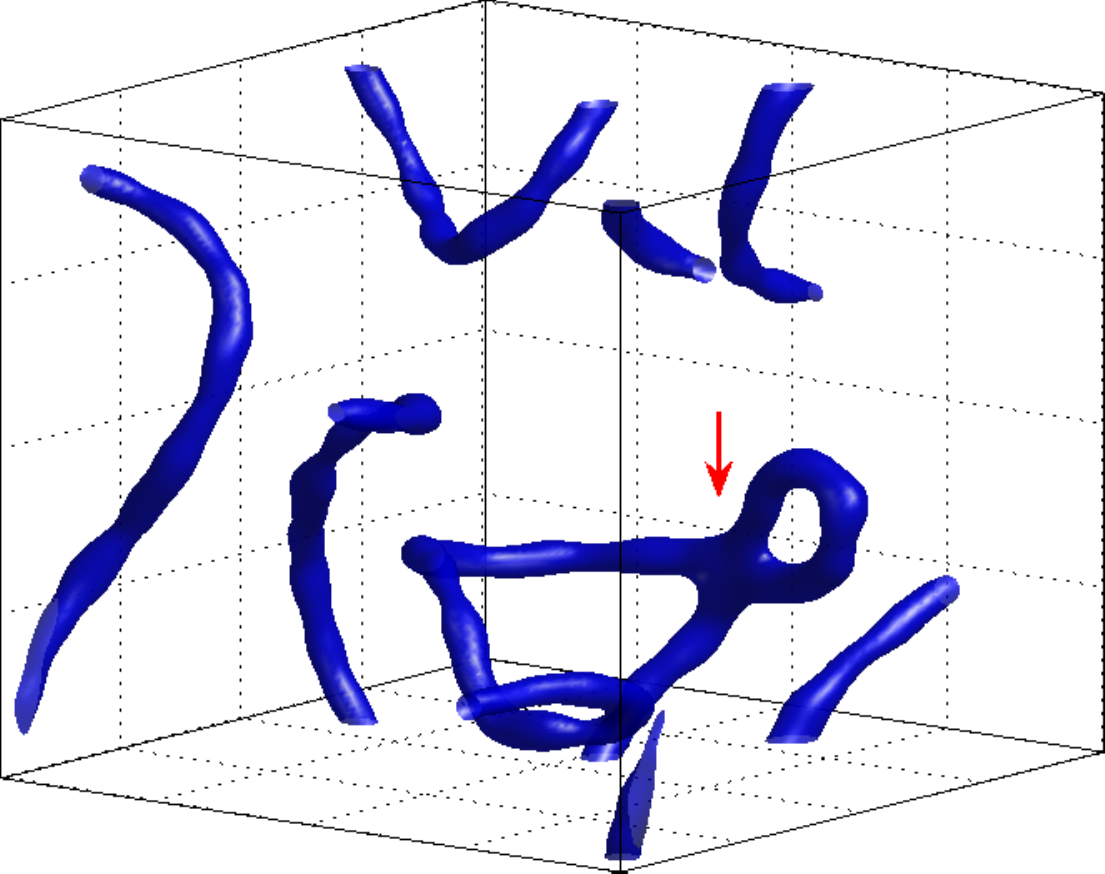}
	\includegraphics[width=4.0cm]{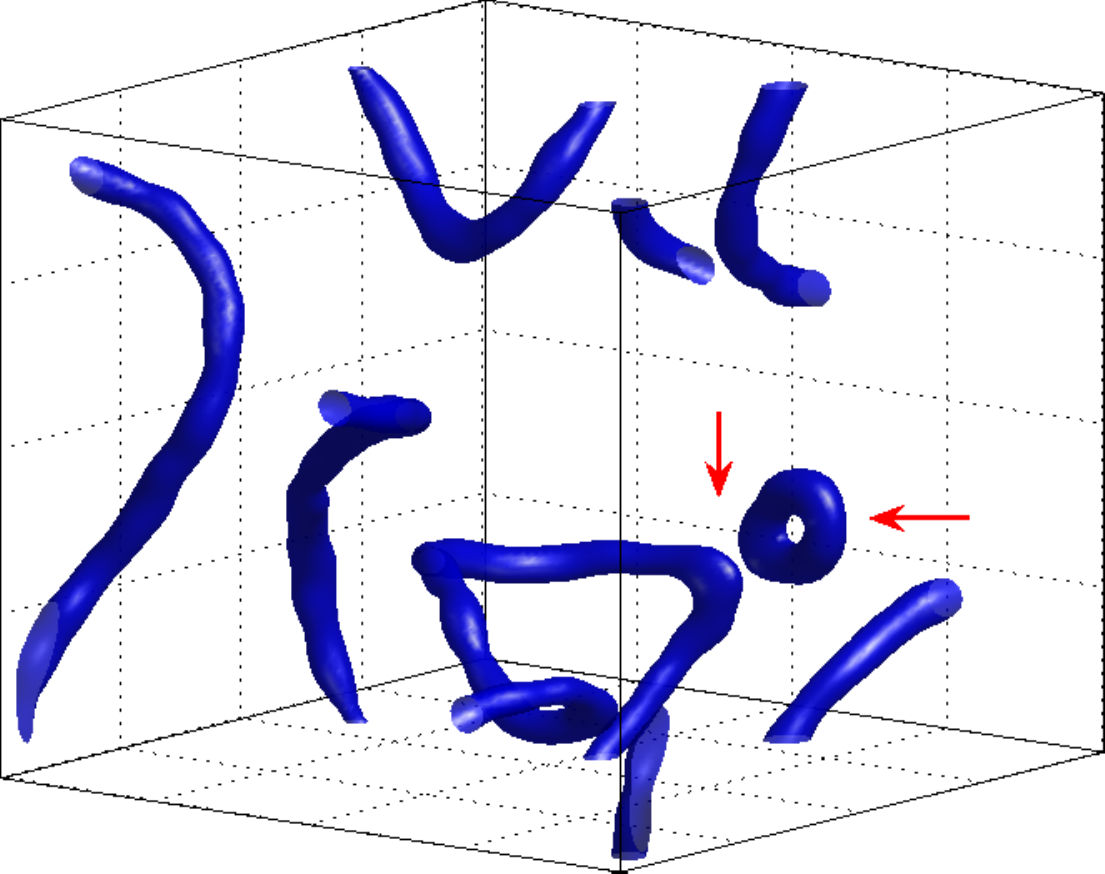}
	\includegraphics[width=4.0cm]{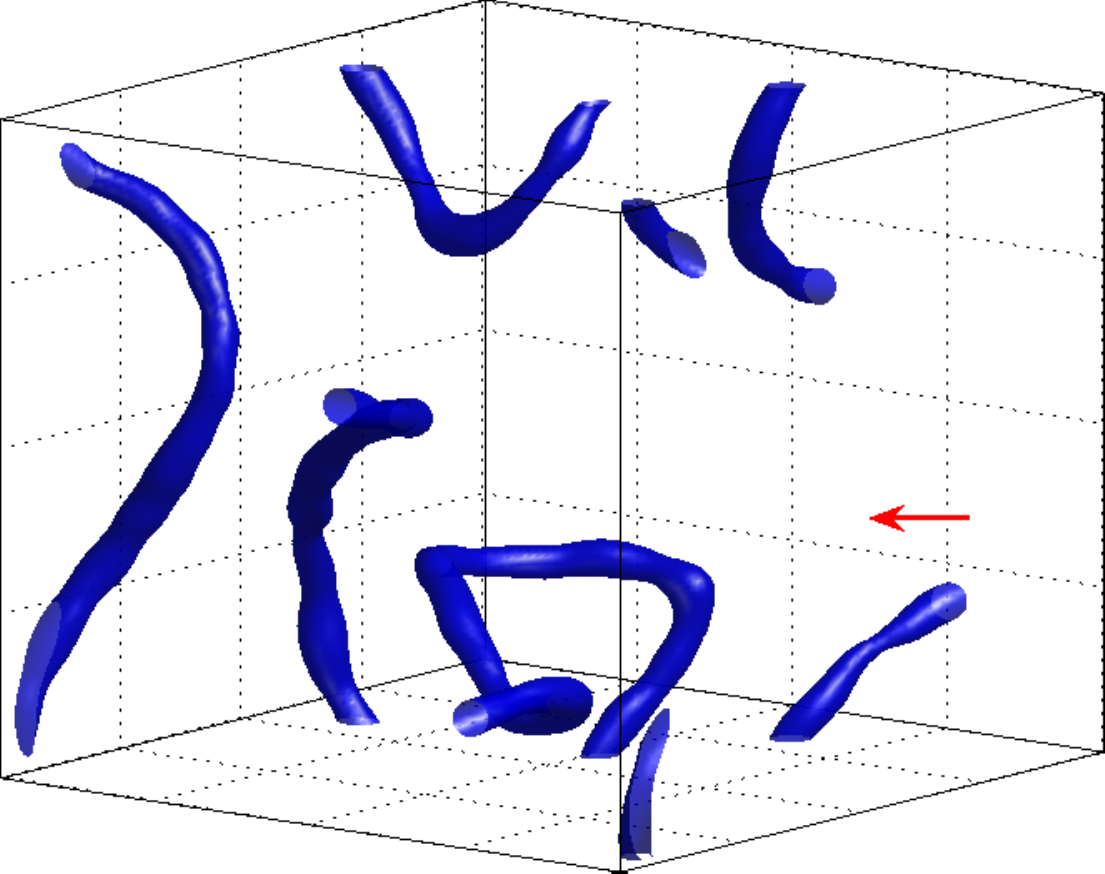}
	\caption{Snapshots obtained after $730$, $850$, $920$ and $1000$
generations of a $128^3$ stochastic network simulation with $N=5$. The
five species are left uncolored while the blue regions represent regions
with a significantly larger density of empty spaces. The vertical arrows
signal an intercommutation process while the horizontal ones indicate
the collapse of a string loop.}
	\label{fig-4} 
\end{figure}

In general, the extension to three spatial dimensions of the dynamics
presented in Fig. \ref{fig-3} gives rises to a curved string network as
shown in Fig. \ref{fig-4}. The snapshots were taken from a $128^3$
simulation after $730$, $850$, $920$ and $1000$ generations. Here the
strings represent regions with a significant larger density of empty
spaces (in order to improve the visualization the density of empty spaces 
has been convolved with a gaussian filter function).

Whenever two strings intersect they
intercommute (exchange partners). This process is responsible for the
production of string loops which collapse with a characteristic velocity
roughly proportional to the loop characteristic scale. The collapse of a
string loop is curvature driven and it is associated to the existence of
a defect/anti-defect pair on any plane intersecting the loop. The
processes of intercommutation and loop collapse are indicated by red
arrows in Fig. \ref{fig-4}.
%%%%%%%%%%%%%%%%%%%%%%%%%%%%%%%%%%%%%%%%%%%%%%%%%%%%%%%%%%%%
%%%%%%%%%%%%%%%%%%%%%%%%%%%%%%%%%%%%%%%%%%%%%%%%%%%%%%%%%%%%

Let us now consider $N+1$ scalar fields ($\phi_0$, $\phi_1$, $\phi_2$,
$\ldots$, $\phi_N$) representing the fraction of space around a given
point occupied by empty sites ($\phi_0$) and by individuals of the
species $i$ ($\phi_i$), satisfying the constraint $\phi_0+\phi_1+ \ldots
+\phi_{N}=1$. For $N \ge 4$ the mean field equations of motion
\begin{eqnarray}
	{\dot \phi}_{0} &=& D \nabla^2 \phi_0 -
	r \phi_0 \sum_{i=1}^{N}\phi_i + \ p\, \sum_{i=1}^{N}
	\sum_{\alpha=2}^{N-2} \phi_{i}\phi_{i+\alpha},
	\label{1}\\
	{\dot \phi}_{i} &=& D \nabla^2 \phi_i +
	r \phi_0 \phi_i 
	-p\sum_{\alpha=2}^{N-2} \phi_{i}\phi_{i+\alpha},
	\label{2}
\end{eqnarray}
describe the average dynamics of the models studied in the present
letter. A dot represents a derivative with respect to time and $D=2m$ is
the diffusion rate.

\begin{figure}[h]
	\centering
	\includegraphics[width=4.0cm]{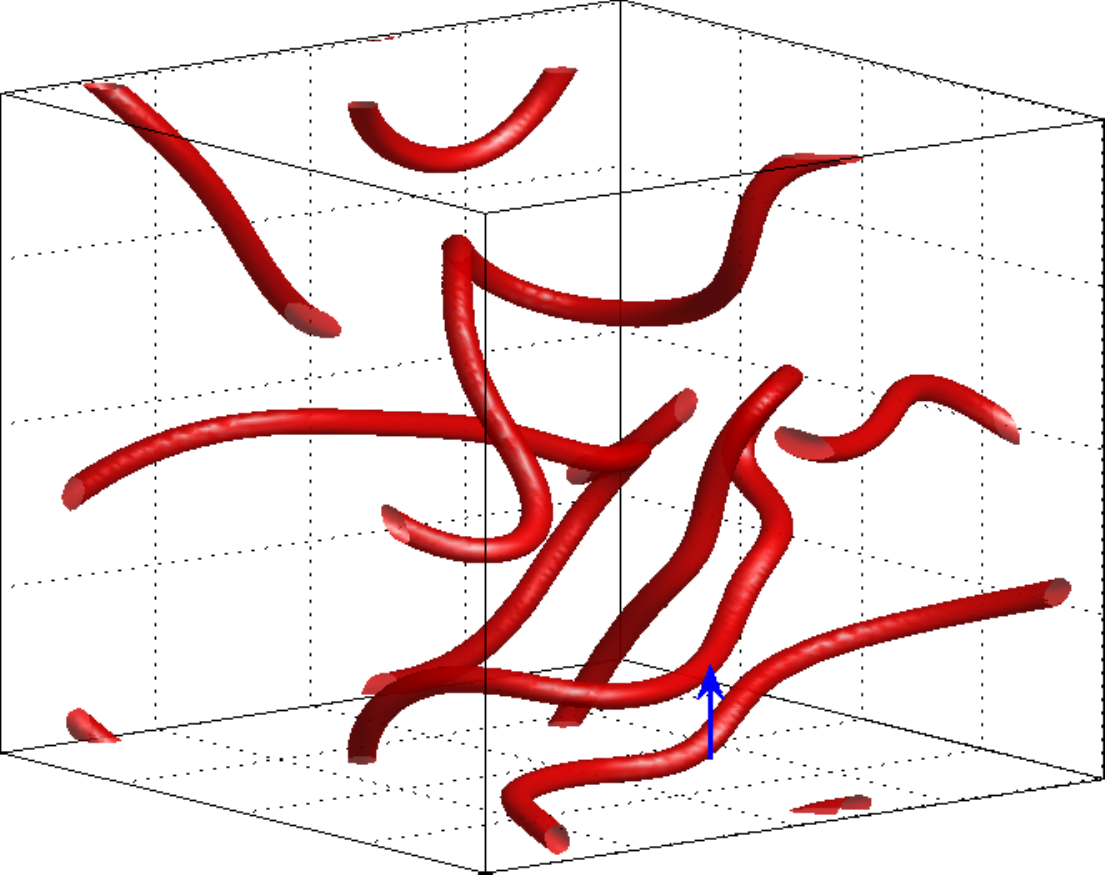}
	\includegraphics[width=4.0cm]{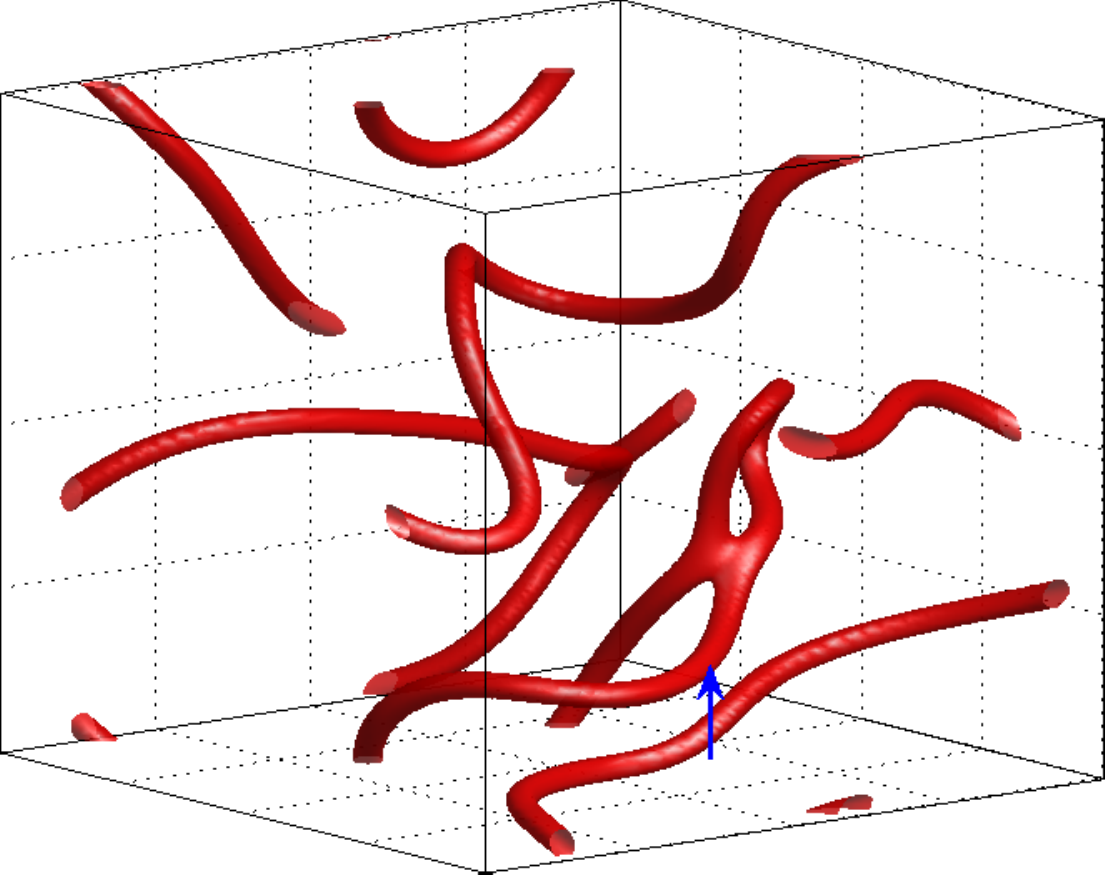}
	\includegraphics[width=4.0cm]{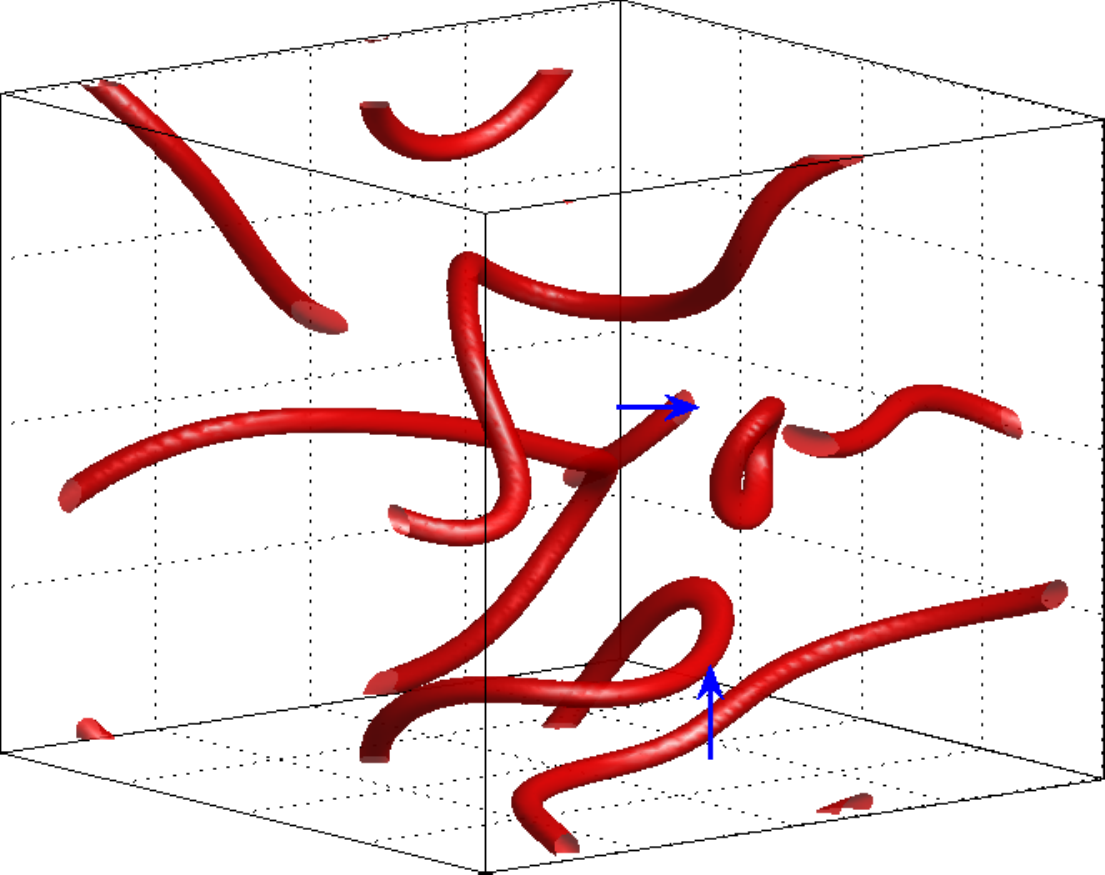}
	\includegraphics[width=4.0cm]{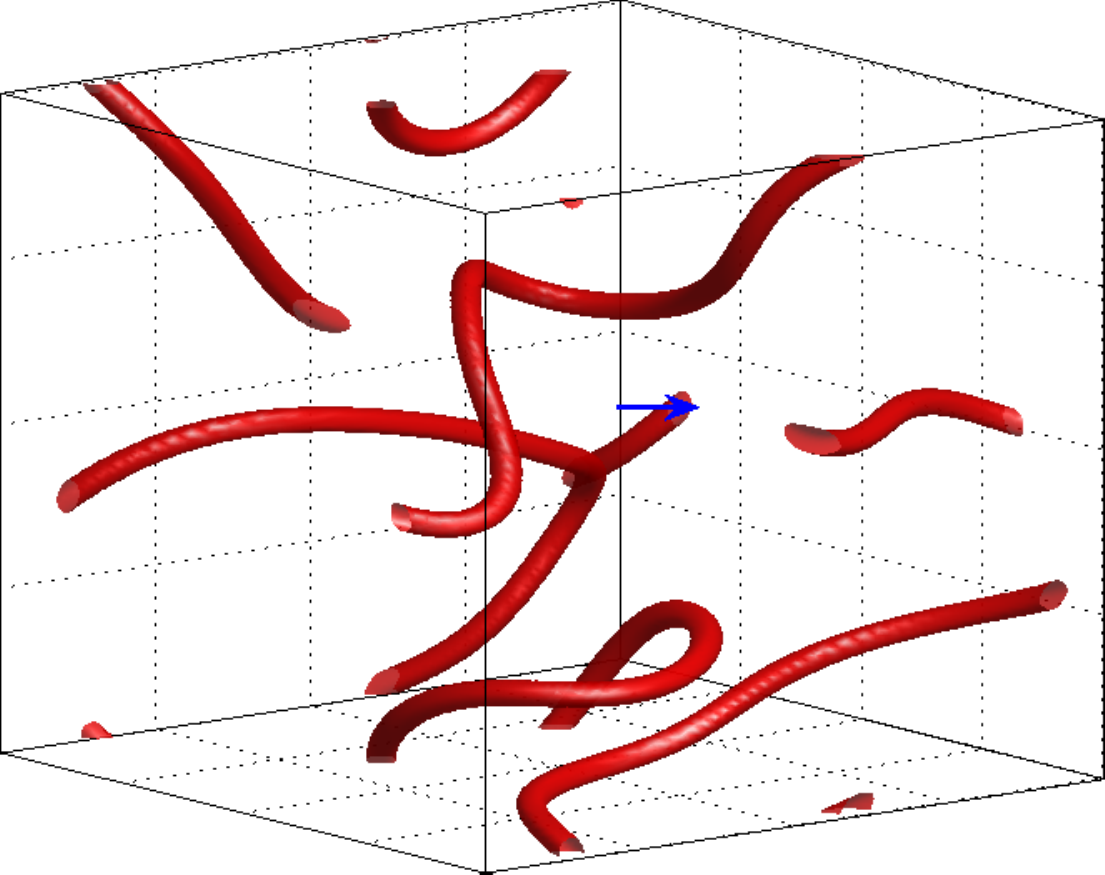}
	\caption{Similar to Fig. \ref{fig-4}, but now the snapshots were
obtained from a mean field simulation of the 4 species model, after
$1120$, $1200$, $1280$ and $1360$ generations.}
	\label{fig-5} 
\end{figure}

We performed a set of three-dimensional mean field network simulations
starting with initial conditions where at each grid point a species $s$
was chosen at random. Initial conditions with $\phi_i=1$ if $i=s$ and
$\phi_i=0$ if $i \neq s$ were set at each grid point ($\phi_0$ was set
to zero at every grid point). Snapshots of a three-dimensional numerical
simulation of the $4$ species model running with $D=0.30$ are shown in
Fig. \ref{fig-5}. One may observe the presence of intercommutations and
the collapse of string loops.
%%%%%%%%%%%%%%%%%%%%%%%%%%%%%%%%%%%%%%%%%%%%%%%%%%%%%%%%%%%%
%%%%%%%%%%%%%%%%%%%%%%%%%%%%%%%%%%%%%%%%%%%%%%%%%%%%%%%%%%%%

Indeed, the results provided by the mean field simulations mirror those
obtained from the stochastic networks, except for the noise (see the 
movie \cite{sync-3D-5S}, done for the $5$ species model, for more 
details).

\begin{figure}[h]
	\centering
	\includegraphics[width=8.0cm]{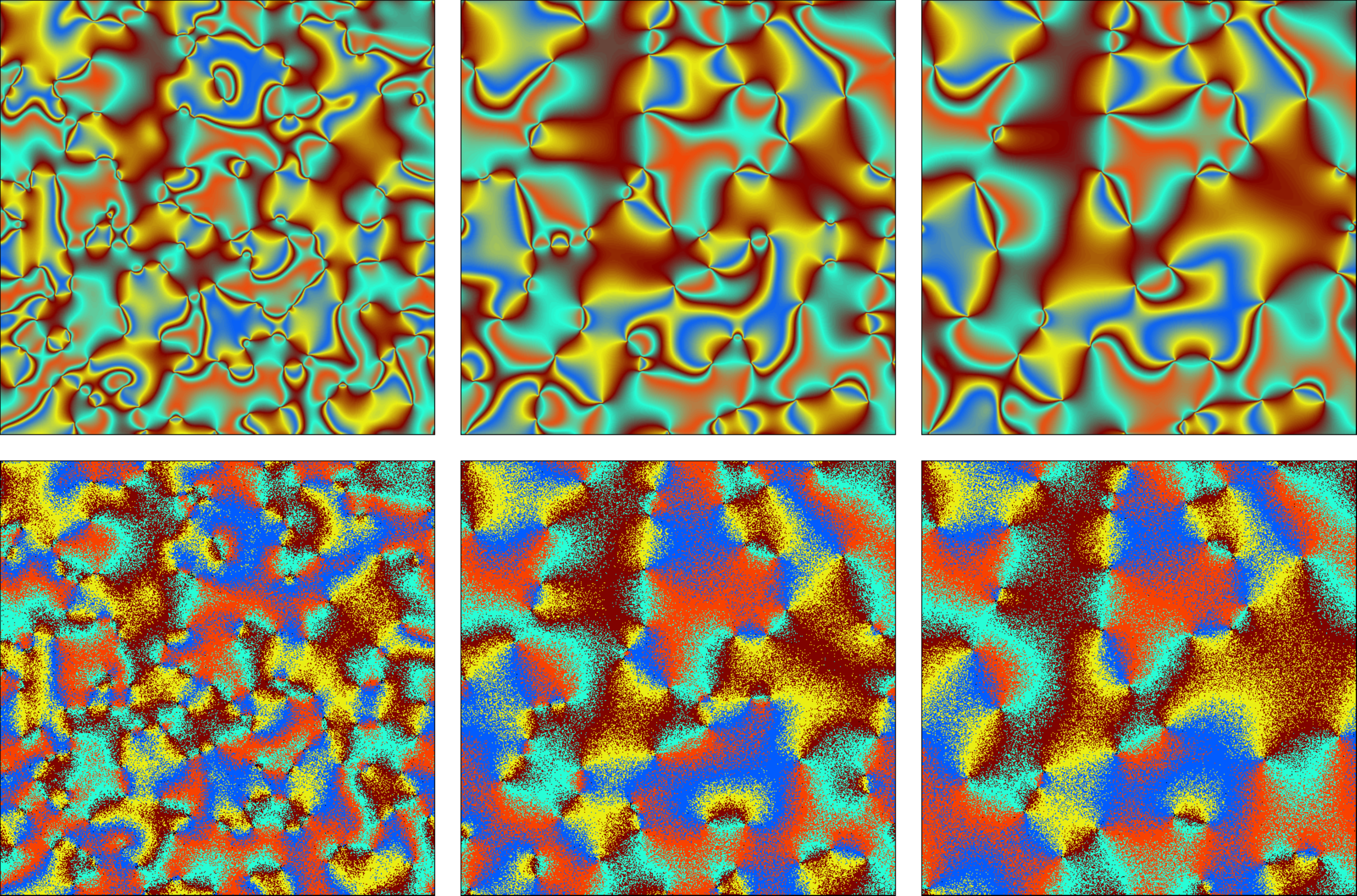}
\caption{Snapshots obtained after $2000$, $4000$ and $6000$ generations
(from left to right, respectively) of $1024^2$ mean field (upper panels)
and stochastic (lower panels) string network simulations of the 5
species model, starting with similar initial conditions.}
 \label{fig-6}
\end{figure}

This correspondence was also verified by comparing snapshots of $1024^2$
simulations starting with similar initial conditions. For this purpose
we started by running the mean field simulations for 2000 generations.
We then constructed the initial conditions of the stochastic simulations
by randomly selecting a species or an empty space, at each grid point,
according to the probability distribution given by the mean field
simulations. Fig. \ref{fig-6} shows three snapshopts from mean-field
(upper panels) and stochastic (lower panels) simulations of the 5 species 
model with similar initial conditions, clearly showing a synchronous evolution. 
Note that the defect properties and dynamics are naturally associated to the coexistence 
between competing species in and around the cores (see Figs. \ref{fig-4} and 
\ref{fig-5} and the movie \cite{sync-3D-5S} for the corresponding dynamics 
in three spatial dimensions).

%%%%%%%%%%%%%%%%%%%%%%%%%%%%%%%%%%%%%%%%%%%%%%%%%%%%%%%%%%%%
%%%%%%%%%%%%%%%%%%%%%%%%%%%%%%%%%%%%%%%%%%%%%%%%%%%%%%%%%%%%

%%%%%%%%%%%%%%%%%%%%%%%%%%%%%%%%%%%%%%%%%%%%%%%%%%%%%%%%%%%%
\begin{figure}[h]
	\centering
	\includegraphics[width=5.0cm]{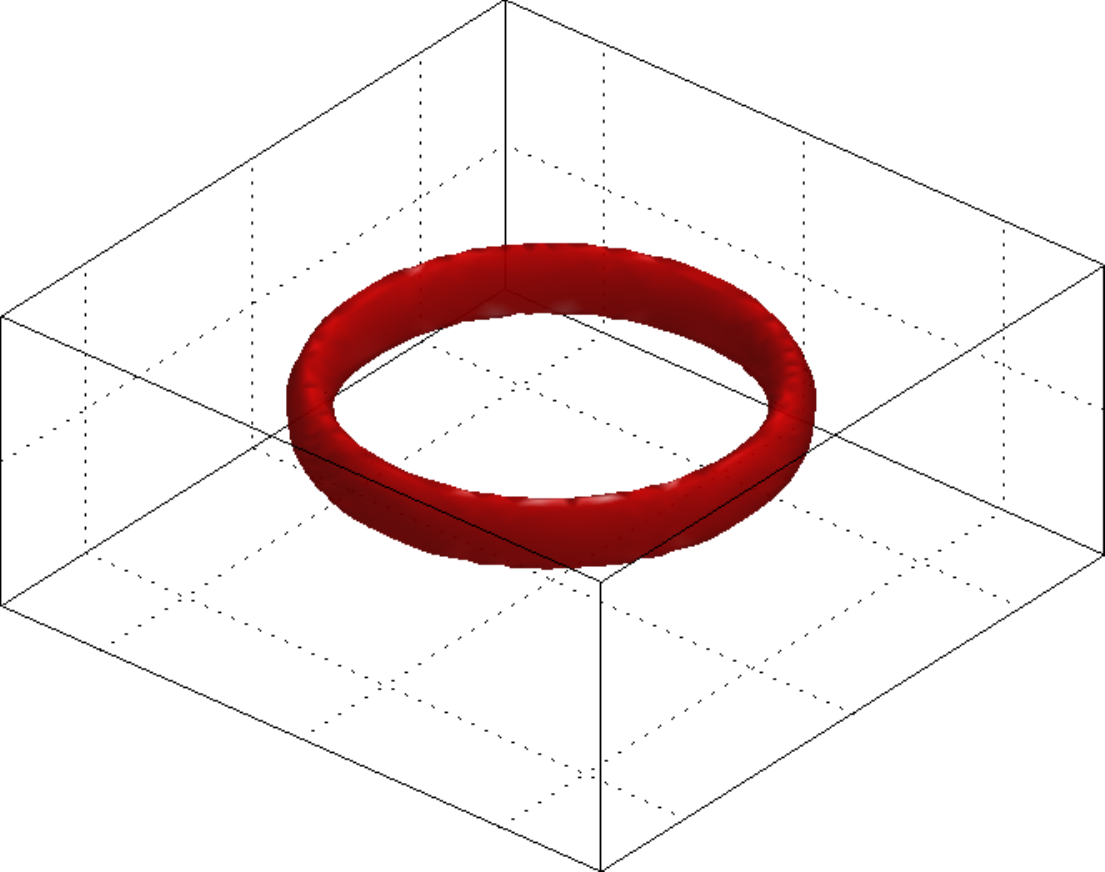}
	\caption{Circular string loop in a model with 5 species.}
	\label{fig-7}
\end{figure}
%%%%%%%%%%%%%%%%%%%%%%%%%%%%%%%%%%%%%%%%%%%%%%%%%%%%%%%%%%%%

\begin{figure}[h]
	\centering
	\includegraphics[scale= 1.0]{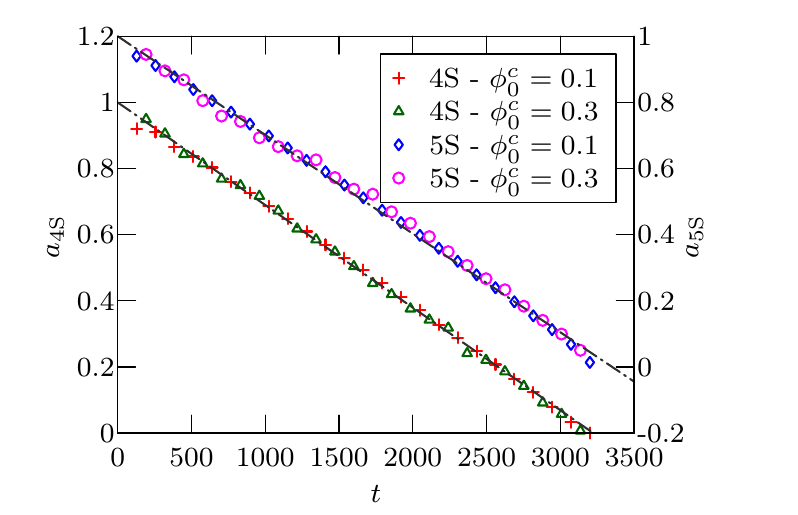}
	\caption{The time evolution of the area of circular strings for models
$N=4$ and $N=5$ with the dashed lines representing the best fits.}
	\label{fig-8}
\end{figure}

In order to better understand the dynamics of string networks in the
context of the spatial stochastic models studied in this letter, we
also consider the collapse of a circular string loop (see Fig.
\ref{fig-7}). The average number of empty spaces per unit string length
($\mu$) does not change significantly with time and, consequently, the
loop perimeter is approximately proportional to the total number of
empty sites associated with it. Therefore the evolution of the area $a$
of the circle limited by the circular string loop can be measured by
determining the variation of the number of empty spaces associated with
the loop. For this purpose we start by defining $\varphi_0({\bf
r},t)\equiv {\rm max}\left(\phi_0({\bf r},t)-\phi_{0}^{c},0\right)$,
where $\phi_{0}^{c}$ represents a threshold which guarantees that only
grid points with a high number density of empty sites, close to the core
of the string, are accounted for. The average number density of empty
sites associated with the string is given by $\rho(t) = {\mathcal
N}^{-2}\displaystyle\sum_{{\bf r}}\,\varphi_0({\bf r},t)$,
%\begin{equation}
%	\rho(t) = \displaystyle\sum_{{\bf r}}\,\varphi_0({\bf r},t)\,,
%\end{equation}
with $a \propto \rho^2$. The time evolution of $a$ for circular loops in
models with 4 and 5 species are shown in Fig. \ref{fig-8} for
$\phi_{0}^{c}=0.1$ and $\phi_{0}^{c}=0.3$. Note that both results
$a_{4S}$ ($N=4$) and $a_{5S}$ ($N=5$) are only weakly dependent on the
threshold and agree with the theoretical relation $a(t) =
a_0\,\left(1-t\big/t_{c}\right)$, where $a_0$ is the initial area and
$t_{c}$ is the collapse time ($a_0$ is normalized to unity at the
initial time $t=0$). This evolution is analogous to that of the area of
the spherical interfaces studied in Ref. \cite{PhysRevE.86.031119}.
%%%%%%%%%%%%%%%%%%%%%%%%%%%%%%%%%%%%%%%%%%%%%%%%%%%%%%%%%%%%
%%%%%%%%%%%%%%%%%%%%%%%%%%%%%%%%%%%%%%%%%%%%%%%%%%%%%%%%%%%%

\begin{figure}[h]
	\centering
	\includegraphics[scale= 1.0]{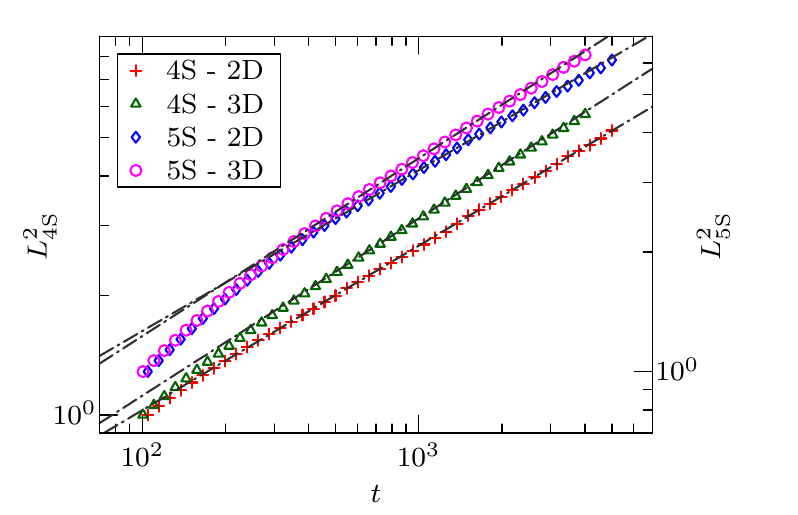}
	\caption{The average scaling exponent $\lambda$ computed from ensebles
of ten $1024^2$ and $256^3$ network simulations of models with $N=4$ and
$N=5$. The best fits are represented by the dashed lines.}
	\label{fig-9}
\end{figure}

The characteristic length of a string network can be defined as $L
\equiv (\mu/\rho)^{1/2} \propto \rho^{-1/2}$. The average evolution of
$L$ with time $t$ was calculated by carrying out sets of $10$ distinct
two- and three-dimensional mean field network simulations ($1024^2$ and
$256^3$) with different initial conditions. Fig. \ref{fig-9} shows that
the characteristic lengths $L_{4S}$ (4 species) and $L_{5S}$ (5 species)
evolve in reasonable agreement with the scaling law $L\propto
t^{\lambda}$, with $\lambda = 1/2$ (in all cases the value of $L$ was
normalized to unity at $t=100$). The results obtained by fiiting a $L\propto
t^{\lambda}$ to the data for $t > 200$ were $\lambda=0.41 \pm 0.01$ ($N=4,2D$), 
$\lambda=0.41 \pm 0.02$ ($N=5,2D$), $\lambda=0.45 \pm 0.02$ ($N=4,3D$), 
$\lambda=0.45 \pm 0.02$ ($N=5,3D$). Note that $\lambda=0.5$ would only be expected 
in an ideal case of simulations with an infinite dynamical resolution and dynamical range. 
Given the limited resolution and dynamical range of the simulations, the exponent 
calculated from the simulations appears to be in reasonable agreement with the 
theoretical expectation (see \cite{Avelino:2010qf, Avelino:2011ev, Sousa:2011ew, 
Sousa:2011iu} for a detailed account of the connection between the $L \propto t^{1/2}$ 
scaling law for the dynamics of domain walls and strings and mean field equations of 
motion of the form ${\dot \phi}=f(\phi)$, where $\phi$ is a scalar field multiplet).

The scaling law $L\propto t^{1/2}$ also describes the dynamics of non-relativistic string networks in
condensed matter and cosmology \cite{PhysRevD.53.R575}, in a friction dominated regime. The characteristic velocity of these string networks is given by $v=L/t \propto t^{-1/2} \propto L^{-1}$, thus leading to a
string network evolution which slows down at a rate which is proportional to the increase of the average radius of curvature of the strings.

%%%%%%%%%%%%%%%%%%%%%%%%%%%%%%%%%%%%%%%%%%%%%%%%%%%%%%%%%%%%
%%%%%%%%%%%%%%%%%%%%%%%%%%%%%%%%%%%%%%%%%%%%%%%%%%%%%%%%%%%%

\section{Conclusions}
\label{conclusions}

The investigation presented in this paper represents a significant
extension with respect to previous works, which have mainly been focused on the 
dynamics of two-dimensional population domains. To the best of our knowledge,
this is the first time that string networks were shown to arise in the
context of generalized Lotka-Volterra competition models. Although string 
networks have been studied in detail in condensed matter and cosmology, we are 
not aware of any previous three-dimensional studies/experiments on biological 
populations leading to string networks. We have performed two- and three-dimensional 
stochastic and mean field theory 
simulations, showing that the coarsening dynamics of these string networks appears 
to follow a scale-invariant evolution. A similar behavior is also found in other physical 
systems, in particular in the case of the curvature driven dynamics of string networks in 
condensed matter and cosmology.

\section*{Acknowledgements}
\label{thanks}

We thank FCT-Portugal, FAPESP, CAPES/Nanobiotec and CNPQ/Fapern for
financial support. P.P.A. is supported by a Investigador FCT contract 
funded by FCT/MCTES (Portugal) and POPH/FSE (EC).



%% References


\bibliographystyle{elsarticle-num}
\bibliography{strings-pla}

\begin{thebibliography}{10}
\expandafter\ifx\csname url\endcsname\relax
  \def\url#1{\texttt{#1}}\fi
\expandafter\ifx\csname urlprefix\endcsname\relax\def\urlprefix{URL }\fi
\expandafter\ifx\csname href\endcsname\relax
  \def\href#1#2{#2} \def\path#1{#1}\fi

\bibitem{May-Leonard}
R.~May, W.~Leonard, Nonlinear aspects of competition between three species,
  SIAM Journal on Applied Mathematics 29 (1975) 243.

\bibitem{Kerr2002}
B.~Kerr, M.~A. Riley, M.~W. Feldman, B.~J.~M. Bohannan, {Mobility promotes and
  jeopardizes biodiversity in rock–paper–scissors games}, Nature 418 (2002)
  171.

\bibitem{Reichenbach2007}
T.~Reichenbach, M.~Mobilia, E.~Frey, {Mobility promotes and jeopardizes
  biodiversity in rock–paper–scissors games}, Nature 448 (2007) 1046.

\bibitem{Volterra}
V.~Volterra, Lecons sur la Theorie Mathematique de la Lutte pour la Vie,
  $1^\mathrm{st}$ Edition, Gauthier-Villars, Paris, (1931).

\bibitem{doi:10.1021/ja01453a010}
A.~J. Lotka, Undamped oscillations derived from the law of mass action.,
  Journal of the American Chemical Society 42~(8) (1920) 1595--1599.

\bibitem{Szabo2008}
G.~Szabo, A.~Szolnoki, I.~Borsos, Self-organizing patterns maintained by
  competing associations in a six-species predator-prey model, Phys. Rev. E 77
  (2008) 041919.

\bibitem{Peltomaki2008}
M.~Peltomaki, M.~Alava, Three- and four-state rock-paper-scissors games with
  diffusion, Phys. Rev. E 78 (2008) 031906.

\bibitem{Durrett:2009}
R.~Durrett, {Coexistence in stochastic spatial models}, The Annals of Applied
  Probability 19 (2019) 477--496.

\bibitem{PhysRevE.81.046113}
W.-X. Wang, Y.-C. Lai, C.~Grebogi, Effect of epidemic spreading on species
  coexistence in spatial rock-paper-scissors games, Phys. Rev. E 81 (2010)
  046113.

\bibitem{Edwards:2010}
K.~F. Edwards, S.~J. Schreiber, Preemption of space can lead to intransitive
  coexistence of competitors, Oikos 119 (2010) 1201--1209.

\bibitem{PhysRevE.85.061924}
J.~Juul, K.~Sneppen, J.~Mathiesen, Clonal selection prevents tragedy of the
  commons when neighbors compete in a rock-paper-scissors game, Phys. Rev. E 85
  (2012) 061924.

\bibitem{Kang20132652}
Y.~Kang, Q.~Pan, X.~Wang, M.~He, A golden point rule in
  rock-paper-scissors-lizard-spock game, Physica A 392 (2013) 2652.

\bibitem{Lutz2013286}
A.~F. L\"{u}tz, S.~Risau-Gusman, J.~J. Arenzon, Intransitivity and coexistence
  in four species cyclic games, Journal of Theoretical Biology 317 (2013)
  286--292.

\bibitem{PhysRevLett.99.238105}
T.~Reichenbach, M.~Mobilia, E.~Frey, Noise and correlations in a spatial
  population model with cyclic competition, Phys. Rev. Lett. 99 (2007) 238105.

\bibitem{PhysRevE.83.011917}
W.-X. Wang, X.~Ni, Y.-C. Lai, C.~Grebogi, Pattern formation, synchronization,
  and outbreak of biodiversity in cyclically competing games, Phys. Rev. E 83
  (2011) 011917.

\bibitem{May-Leonard-Mobilia}
Q.~He, M.~Mobilia, U.~T\"{a}uber, Coexistence in the two-dimensional
  may-leonard model with random rates, The European Physical Journal B 82
  (2011) 97--105.

\bibitem{PhysRevE.86.021911}
D.~Lamouroux, S.~Eule, T.~Geisel, J.~Nagler, Discriminating the effects of
  spatial extent and population size in cyclic competition among species, Phys.
  Rev. E 86 (2012) 021911.

\bibitem{PhysRevE.86.031119}
P.~P. Avelino, D.~Bazeia, L.~Losano, J.~Menezes, von Neummann's and related
  scaling laws in rock-paper-scissors-type games, Phys. Rev. E 86 (2012)
  031119.

\bibitem{PhysRevE.86.036112}
P.~P. Avelino, D.~Bazeia, L.~Losano, J.~Menezes, B.~F. Oliveira, Junctions and
  spiral patterns in generalized rock-paper-scissors models, Phys. Rev. E 86
  (2012) 036112.

\bibitem{PhysRevE.87.032148}
A.~Roman, D.~Dasgupta, M.~Pleimling, Interplay between partnership formation
  and competition in generalized may-leonard games, Phys. Rev. E 87 (2013)
  032148.

\bibitem{1742-5468-2012-07-P07014}
A.~Roman, D.~Konrad, M.~Pleimling, Cyclic competition of four species: domains
  and interfaces, Journal of Statistical Mechanics: Theory and Experiment
  (2012) P07014.

\bibitem{Jiang20122292}
L.-L. Jiang, W.-X. Wang, Y.-C. Lai, X.~Ni, Multi-armed spirals and multi-pairs
  antispirals in spatial rock–paper–scissors games, Physics Letters A 376
  (2012) 2292--2297.

\bibitem{Avelino2005}
P.~P. Avelino, J.~C. R.~E. Oliveira, C.~J. A.~P. Martins, Understanding domain
  wall network evolution, Phys. Lett. B 610 (2005) 1--8.

\bibitem{Avelino2008}
P.~P. Avelino, C.~J. A.~P. Martins, J.~Menezes, R.~Menezes, J.~C. R.~E.
  Oliveira, Dynamics of domain wall networks with junctions, Phys. Rev. D 78
  (2008) 103508.

\bibitem{PhysRevD.79.085007}
P.~P. Avelino, D.~Bazeia, R.~Menezes, J.~C. R.~E. Oliveira, Bifurcation and
  pattern changing with two real scalar fields, Phys. Rev. D 79 (2009) 085007.

\bibitem{Stavans1989}
J.~Stavans, J.~A. Glazier, Soap froth revisited: Dynamic scaling in the
  two-dimensional froth, Phys. Rev. Lett. 62 (1989) 1318--1321.

\bibitem{Glazier1992}
J.~A. Glazier, D.~Weaire, {The kinetics of cellular patterns}, J. Phys. Cond.
  Mat. 4 (1992) 1867--1894.

\bibitem{Flyvbjerg1993}
H.~Flyvbjerg, Model for coarsening froths and foams, Phys. Rev. E 47 (1993)
  4037--4054.

\bibitem{Monnereau1998}
C.~{Monnereau}, M.~{Vignes-Adler}, {Dynamics of 3D Real Foam Coarsening}, Phys.
  Rev. Lett. 80 (1998) 5228--5231.

\bibitem{Weaire2000}
D.~Weaire, R.~Hutzler, The physics of foams, Oxford University Press, Oxford,
  2000.

\bibitem{PhysRevE.74.061605}
S.~G. Kim, D.~I. Kim, W.~T. Kim, Y.~B. Park, Computer simulations of
  two-dimensional and three-dimensional ideal grain growth, Phys. Rev. E 74
  (2006) 061605.

\bibitem{Avelino:2010qf}
P.~Avelino, R.~Menezes, J.~Oliveira, {Unified paradigm for interface dynamics},
  Phys.Rev. E83 (2011) 011602.

\bibitem{Avelino:2011ev}
P.~P. Avelino, L.~Sousa, {Domain wall network evolution in (N+1)-dimensional
  FRW universes}, Phys.Rev. D83 (2011) 043530.

\bibitem{Sousa:2011ew}
L.~Sousa, P.~P. Avelino, {p-brane dynamics in (N+1)-dimensional FRW universes:
  a unified framework}, Phys.Rev. D83 (2011) 103507.

\bibitem{Sousa:2011iu}
L.~Sousa, P.~P. Avelino, {The cosmological evolution of p-brane networks},
  Phys.Rev. D84 (2011) 063502.

\bibitem{PhysRevD.53.R575}
C.~J. A.~P. Martins, E.~P.~S. Shellard, Scale-invariant string evolution with
  friction, Phys. Rev. D 53 (1996) R575--R579.

\bibitem{Vilenkin:2000}
A.~{Vilenkin}, E.~P.~S. {Shellard}, Cosmic Strings and Other Topological
  Defects, Cambridge University Press, 2000.

\bibitem{Oliveira:2010}
B.~F. {de Oliveira}, P.~P. {Avelino}, F.~{Moraes}, J.~C.~R.~E. {Oliveira},
  {Nematic liquid crystal dynamics under applied electric fields}, Phys. Rev. E
  82 (2010) 041707.

\bibitem{sync-3D-5S}
\href{http://www.youtube.com/watch?v=l9MinDd7AzQ}{[link]}.
\newline\urlprefix\url{http://www.youtube.com/watch?v=l9MinDd7AzQ}

\end{thebibliography}






% \end{thebibliography}

\end{document}